\pdfoutput=1
\documentclass[12pt]{article}
\usepackage{graphicx,amssymb,lipsum,mathtools,amsmath,amssymb}

\def\ea{\end{array}}

\def\bc{\begin{center}}
\def\ec{\end{center}}
\usepackage{graphicx}
\usepackage{xcolor}
\usepackage{fancyhdr}
\usepackage{ae,aecompl}
\usepackage{epsfig}
\usepackage{mathtools}
\usepackage{ mathrsfs }
\usepackage{ upgreek }
\usepackage{ dsfont }
\usepackage{IEEEtrantools}
\usepackage{verbatim}
\usepackage[english]{babel}

\usepackage{amsfonts}
\usepackage{amssymb}
\usepackage{lscape}
\usepackage{amsmath}
\usepackage{textcomp}
\usepackage{txfonts}   
\usepackage{pstricks}
\usepackage{wrapfig}
\usepackage{multirow}
\usepackage{booktabs}
\usepackage{textcomp}

\begin{document}

\title{Studying the physics potential of long-baseline experiments in terms of new sensitivity
parameters }
\author{Mandip Singh$^*$ \\
\\
{ \it Department of Physics, Centre of Advanced Study, P. U., Chandigarh, India.}\\
}

\maketitle
\begin{abstract}
We investigate physics opportunities to constraint leptonic CP-violation phase $\delta_{CP}$
through numerical analysis of working neutrino oscillation probability parameters, in the context of
long base line experiments. Numerical analysis of two parameters, the `` transition probability 
$\delta_{CP}$ phase sensitivity parameter ($A^M$) '' and `` CP-violation probability $\delta_{CP}$ 
phase sensitivity parameter ($A^{CP}$) '', as function of beam energy and/or base line has been 
preferably carried out. It is an elegant technique to broadly analyze different 
experiments to constraint $\delta_{CP}$ phase and 
also to investigate mass hierarchy in the leptonic sector. 
The positive and negative values of parameter $A^{CP}$ corresponding 
to either of hierarchy in the specific beam energy ranges, could be a very 
promising way to explore mass hierarchy and $\delta_{CP}$ phase. The keys to 
more robust bounds on $\delta_{CP}$ phase are improvements of the involved detection techniques 
to explore bit low energy and relatively long base line regions with better experimental accuracy.
\end{abstract}

\newpage
\maketitle
                                               \section{Introduction}
\label{section:introductio}
Phenomenon of neutrino oscillations in vacuum and matter can be described by six fundamental parameters:
three lepton flavor mixing angles viz. $\theta_{12}$; $\theta_{13}$; $\theta_{23}$, two neutrino mass--squared differences
$\Delta m_{21}^2$; $\Delta m_{23}^2$ and one Dirac-type CP-violating phase $\delta_{CP}$, collectively known as 
neutrino oscillation parameters. Owing to a number of dedicated neutrino oscillation experiments in the past
decades, both ($\theta_{12}^2$; $\Delta m^2_{21}$) and ($\theta_{23}$; $|\Delta m^2_{31}|$) 
have been measured with reasonably good accuracy \cite{int1}. The investigation of moderately large value of 
smallest leptonic mixing angle $\theta_{13}$ in the investigation of lepton mixing matrix \cite{int2}, \cite{int3}, \cite{int4}
by the Daya Bay \cite{int5} and RENO \cite{int6} reactor neutrino experiments
has rejuvenated the opportunities to investigate unknowns in the neutrino physics. 
This great discovery enhances the possible capability of the next-generation experiments
to pin down the neutrino mass hierarchy (i.e., the sign of $\Delta m^2_{31}$) and eventually to determine the leptonic
Dirac CP-violating phase $\updelta_{CP}$. 
 Global fit of neutrino oscillations with data from world class experiments \cite{glob1}, \cite{glob2} put 
stringent bounds on the neutrino oscillation parameters. 

In the present work we shall discuss about the possible measurement of CP-violating phase `$\delta_{CP}$', 
in the context of recently proposed LAGUNA-LBNO\cite{lbno} and LBNE\cite{lbne} experiments.

Long Base Line (LBL) neutrino experiments like LBNO, LBNE etc. 
  due to their long base lines have advantage over the short base 
   line experiments, latter can be approximated to vacuum oscillation neutrino experiments.
    In vacuum, CP-violation depends only on $\delta_{CP}$ phase, hence vacuum 
     oscillation CP-violation amplitudes give pure or intrinsic measurement of
      $\delta_{CP}$. 
        Due to very small values of CP-violating effects at these short base lines,
          it is very difficult to carry out their experimental analysis. 
           Over long distances contamination of terrestrial matter effects becomes 
            large, which in turn increases oscillation amplitude and fake the $\delta_{CP}$ phase effects.
 In LBL experiments pure CP-violation effects arising due to $\delta_{CP}$ phase only get mixed with CP-violation matter effects
 arising due to asymmetric forward scattering of neutrino's and anti-neutrino's with matter constituents, 
   also known as fake or extrinsic CP-violation effects. 
   In case of matter oscillation phenomenology, CP conjugate of particle oscillation probability 
   can be obtained by merely changing the sign of $\delta_{CP}$ phase and matter potential `A' 
   (as can be seen in equations (\ref{seri}) and (\ref{antseri}) below).
   Due to these changes, matter effects in the case of normal mass hierarchy produce overall enhancement in 
   the vacuum effects, which makes transition probability amplitude so large at moderate 
       base line lengths that, we expect them to measure experimentally. But now if
        we shift from the normal mass hierarchy (NH i.e. $\Delta m_{13}^{2}>0$)
         to the inverted mass hierarchy (IH i.e. $\Delta m_{13}^{2}<0$),
         the mass hierarchy parameter $\upalpha$ in equation (\ref{seri})
           also changes sign, due to which a part of matter 
            effects get reduced, which in turn lowers the 
            value of probability amplitude.
             This addition in the NH-case
              and subtraction 
              in the IH-case at given base line length `L'
              and beam energy `E', separates the NH and IH probability 
               amplitudes to the amount that we can differentiate among
                them experimentally. 

In LBL experiments, the experimental configurations: LBNE($L=1280 ~km, ~E=3.55 \pm 1.38 ~GeV$) 
and LBNO($L=2300 ~km, ~E=5.05 \pm 1.65 ~GeV$) \cite{tohln}
are so chosen, that the asymmetry 
between $\upnu_{\mu} \rightarrow \nu_{e}$ and $\overline{\nu}_{\mu} \rightarrow \overline{\nu}_{e}$
oscillation probabilities is larger than the CP violation effects produced by $\delta_{CP}$ phase, which makes these suitable
for determining the mass hierarchy as well as $\delta_{CP}$ phase \cite{mocpv}. The recently proposed neutrino oscillation experiment viz.
DUNE \cite{dune1}, \cite{dune2}, \cite{dune3} with base line nearly equal to LBNE, holds similar discussion and conclusions to that of LBNE.
Thus while studying LBNE, we are also studying oscillation phenomenology of the DUNE experiment simultaneously.

                               \section{Oscillation phenomenology of platinum channel}
\label{section:phenomenology}

The sub-dominant platinum channel ($\nu_{\mu} \rightarrow \nu_e$), because of its sensitivity to still unknown 
neutrino oscillation parameters (e.g. mass ordering, $\delta_{CP}$ phase, octant of $\theta_{23}$ etc) 
and ability to analyze experimental data logically, has the advantage over other appearance and disappearance oscillation channels.
The analytic expressions for neutrino flavor transition probabilities up to first and/or 
  second order in small oscillation parameters viz. mass ordering parameter (ratio of the solar to atmospheric mass
   square differences, i.e. $\alpha=\Delta m_{21}^{2}/ \Delta m_{31}^{2}$) and third mixing angle `$\theta_{13}$' 
    (also known as reactor mixing angle) has been already calculated in the literature by \cite{hminkta}, \cite{acrvera},
      \cite{amgago} and \cite{akhmet} very elegantly. All these analytic formalisms make use of the method of perturbation theory
      expansion of neutrino evolution $\mathcal{S}$-matrix. In the present work, we have preferably made use of the 
       platinum channel oscillation probability from analytic results by \cite{akhmet}, that can be written as
 \begin{eqnarray}
    P_{\mu \rightarrow e} \equiv \left|\mathcal{S}^{f}_{e, \mu} \right|^{2} &=& \alpha^{2} ~sin^{2} {2\theta_{12}}~ c^{2}_{23}~ 
    \frac{sin^{2}{[A \Delta 
     \frac{L}{2}]}}{A^{2}} + 4~ s^{2}_{13}~ s^{2}_{23} ~\frac{sin^{2}{[(A-1)\Delta \frac{L}{2}]}}{(A-1)^{2}}  \nonumber \\ 
      && +  ~2 ~\alpha~ s_{13} ~sin {2 \theta_{12}} ~sin {2\theta_{23}}~ cos{(\Delta \frac{L}{2} + \delta_{CP})}~ 
       \frac{sin~{[A \Delta \frac{L}{2}]}}{A} ~ \frac{sin{[(A-1) \Delta} \frac{L}{2}]}{(A-1)}
        \label{seri}
         \end{eqnarray}
An another reason for preferring platinum channel 
   lies in the fact, that now a days charged
     mu-mesons can easily be stored in
      world class facility 
       accelerator beam
        dump sources \cite{wjmar}, \cite{btab}, \cite{btab1}, \cite{nufact}, \cite{nufact1}, 
         which can be controlled to accelerate
          these charged entities to the desired energy values. 

The transition probability for anti-neutrinos can be obtained by merely changing 
   $\delta_{CP} \rightarrow -\delta_{CP}$ and $V \rightarrow -V$(or $A\rightarrow-A$) in equation (\ref{seri}) above,
    hence we can write
\begin{eqnarray}
 P_{\overline{\mu} \rightarrow \overline{e}} &=& \alpha^{2} ~sin^{2} {2\theta_{12}}~ c^{2}_{23}~ \frac{sin^{2}{[A \Delta 
    \frac{L}{2}]}}{A^{2}} + 4~ s^{2}_{13}~ s^{2}_{23} ~\frac{sin^{2}{[(A+1)\Delta \frac{L}{2}]}}{(A+1)^{2}}  \nonumber \\ 
     && +  ~2 ~\alpha~ s_{13} ~sin {2 \theta_{12}} ~sin {2\theta_{23}}~ cos{(\Delta \frac{L}{2} - \delta_{CP})}~ 
      \frac{sin~{[A \Delta \frac{L}{2}]}}{A} ~ \frac{sin{[(A+1) \Delta} \frac{L}{2}]}{(A+1)}
       \label{antseri}
\end{eqnarray}
with $A \equiv 2~E~V/\Delta m_{31}^{2}$, where $V = \sqrt{2} ~G_{F} ~N_{e}$; with $N_{e}$ 
  is the number density of electrons in the medium; $G_{F}$ = Fermi weak coupling 
    constant = $11.6639 \times 10^{-24} ~eV^{-2}$, $\Delta \equiv
     \Delta m_{31}^{2}/2~E \simeq \Delta m_{32}^{2}/2~ E$,
      $\upalpha = \Delta m^{2}_{21}/\Delta m^{2}_{32}$,
      $L$ is base line length 
         and $E$ the beam energy. 
 
Above Eq. (\ref{seri}) can be rewritten to the form:
\begin{subequations}
\begin{eqnarray}
  P_{\mu e} =  a ~ + ~ b ~ + ~ c1 ~ cos{~\delta_{CP}} ~ + ~ c2 ~ sin{~ \delta_{CP}}
  \label{nseri}
\end{eqnarray}
with a and b the first and second terms as in Eq. (\ref{seri}) above,
  these are independent of $\delta_{CP}$ phase and the remaining coefficients c1 
    and c2 of $\delta_{CP}$ dependent terms have the following expressions: 
\begin{eqnarray}
 c1 &=& 2 ~ \alpha ~ s_{13} ~ sin{2 \theta_{12}} ~ sin{2 \theta_{23}} ~ \frac{sin(A \Delta L/2)}{A} \frac{sin[(A-1) \Delta L/2]}{(A-1)} 
 ~cos(\Delta L/ 2) \nonumber \\
 c2 &=& - 2 ~ \alpha ~ s_{13} ~ sin{2 \theta_{12}} ~ sin{2 \theta_{23}} ~ \frac{sin(A \Delta L/2)}{A} \frac{sin[(A-1) \Delta L/2]}{(A-1)} 
   ~sin(\Delta L/ 2)
 \label{coeff}
\end{eqnarray}
%
Eq. (\ref{nseri}) can further be compacted to the following form: 
 \begin{eqnarray}
  P_{\mu e}(\delta_{CP}) =  a ~ + ~ b ~ + ~ \sqrt{c1^{2} ~ + ~ c2^{2}} ~ sin(\beta + \delta_{CP})
  \label{serin}
 \end{eqnarray}
\end{subequations}
where $\beta = tan^{-1}(c1/c2)$. 

We can analogously compact anti-particle probability given in Eq. (\ref{antseri}) to the form similar to the above equation.
%
%
\setlength{\arrayrulewidth}{0.27 mm} 
\setlength{\tabcolsep}{7pt}  
\renewcommand{\arraystretch}{1.3} 
\newcommand{\head}[1]{\textnormal{\textbf{#1}}} 
\begin{table*}[htbp!]
\centering
\caption{The best fit and $3 \sigma$ values of mixing angles and mass square differences
from global fit of neutrino oscillation data, adapted from \cite{mtort}.
\label{tab:bfit}}
\begin{tabular}{ p{3cm} p{3cm} p{3cm} }  \hline
 \head{Parameter} &   \head{best fit} {{\bf \fontsize{11}{31}\selectfont$\pm 1\sigma$}} 
 & {{\bf \fontsize{11}{31}\selectfont 3$\sigma$}}  \\ 
\hline
$\theta_{12}^o$ & $34.6 \pm 1.0$ & 31.8 -- 37.8  \vspace{0.2cm}\\ 
$\theta_{23}^o$[NH] & $48.9^{+1.9}_{-7.4}$ & 38.80 -- 53.30  \\
$\theta_{23}^o$[IH] & $49.2^{+1.5}_{-2.5}$ & 39.40 -- 53.10  \vspace{0.2cm}\\
$\theta_{13}^o$[NH] &  $8.8 \pm 0.4$ &  7.70 -- 9.90  \\
$\theta_{13}^o$[IH] &  $8.9 \pm 0.4$ &  7.80 -- 9.90  \vspace{0.2cm}\\
$\Delta m_{21}^2$ & $7.60^{+0.19}_{-0.18}$ &  7.11 - 8.18  \vspace{0.2cm}\\  
$|\Delta m_{31}^2|_{NH}$ & $2.48^{+0.05}_{−0.07}$  &  2.30 -- 2.65  \\
$|\Delta m_{31}^2|_{IH}$ & $2.38^{+0.05}_{-0.06}$  &  2.20 -- 2.54  \\
\hline
\end{tabular}
\end{table*}
%
                  \section{Transition probability, {{\bf \fontsize{16}{31}\selectfont$\delta_{CP}$}}
                   phase sensitivity parameter {{\bf \fontsize{16}{31}\selectfont  ($A^M$)}}}
                    \label{sec:cpoptimization}

\begin{figure*} 
\caption{(Color online)  The variation in the amplitude of transition probability for platinum channel
as a function of beam energy (E).   Sub-figure on LHS represents the LBNE ($L=1,280$ km, $E=3.55$ $\pm$ $1.38~ GeV$) and that on RHS to
the LAGUNA-LBNO ($L=2,300 ~km$, $E=5.05$ $\pm$ $1.65 ~GeV$) experimental setups, as tabulated in table $\ref{tab:lblcon}$. The green curve 
corresponds to band width (i.e. parameter $A^M$) in NH-case and yellow curve in the IH-case, for the full variation (i.e. $0-2 ~\pi$) in
$\delta_{CP}$ phase, as calculated in Eq. (\ref{optpra}). The remaining oscillation parameters have the best fit values shown
in table \ref{tab:bfit}.}
\includegraphics[width=1.0\textwidth]{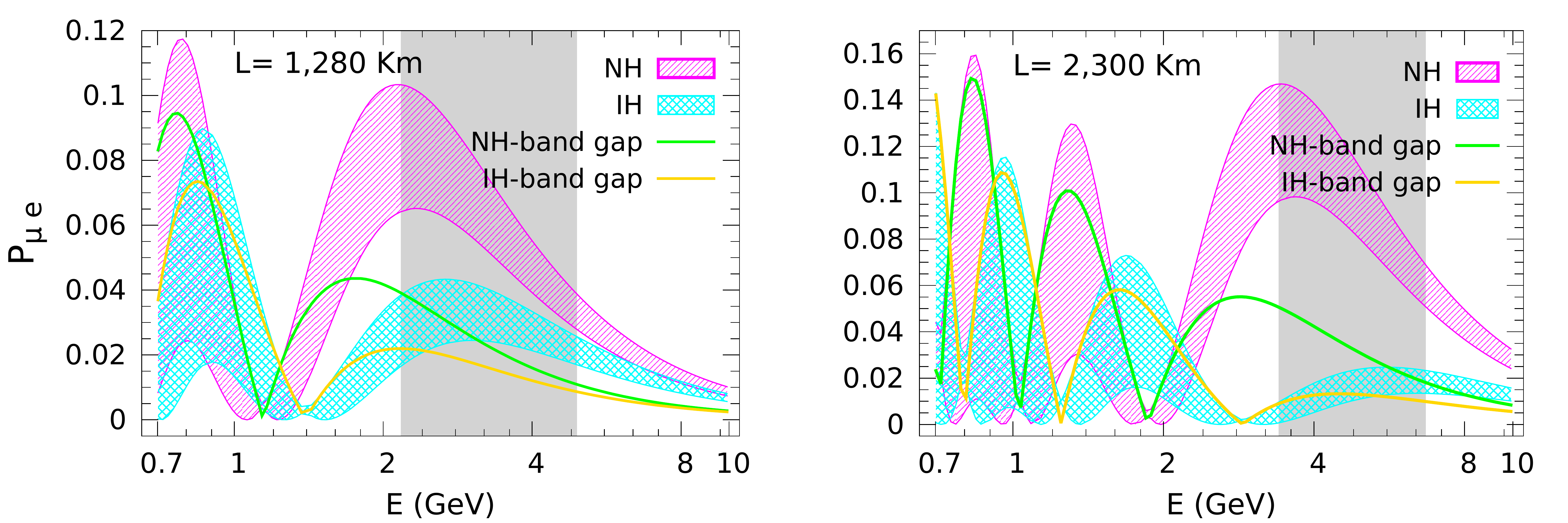}
     \label{fig:optmz}
\end{figure*}

This parameter enables us to predict the sensitivity of the transition probability towards the $\delta_{CP}$ phase 
variations for given experimental configuration.
We can find the maximum possible transition probability amplitude band width ($A^M$)
for full variation in CP-violation phase $\delta_{CP}$ from 0 to 2$\pi$ radians,
  at any chosen value of beam energy $`E'$ and base line $`L'$, with the help of Eq.(\ref{serin}) to the following form 
 \begin{eqnarray}
 \Delta P_{\mu e}^{m}(\delta_{CP}) &\equiv& A^M (say)  \nonumber\\
 &=& P_{\mu e}^{max}(\delta_{CP}) - P_{\mu e}^{min}(\delta_{CP}) 
 = 2 ~\sqrt{c1^{2}+c2^{2}} 
 \label{optpra}
\end{eqnarray}
%
A similar type of parameter has been earlier studied in \cite{dcpm, dcp1, dcp2}.
This parameter is plotted as the green and the yellow colored curves for NH and IH cases respectively in Fig. \ref{fig:optmz}. 
In the NH-case i.e. green colored curve for LBNE, first oscillation maxima of the parameter $A^M$ lies at 1.6 GeV with value 
$\approx 5 \%$ and second maxima at $\approx 0.8$ GeV with value $\approx10 \%$. Similarly for LBNO, first maxima is at 2.8 GeV
with value $\approx 6 \%$ and second is at $\approx 1.3$ GeV with value of $\approx10 \%$. Hence, we can conclude, 
that both experiments are equally sensitive to the variations in $\delta_{CP}$ phase, in the NH-case. 

In the IH-case i.e. yellow colored curve for LBNE, the first and second oscillation maximas exist respectively at 2.2 GeV ($2 \%$) and  0.9 GeV ($7 \%$)
, where values in parentheses are the corresponding values for the parameter $A^M$. Similarly for LBNO the first, second 
and third oscillation maximas are respectively located at 4.2 GeV ($\approx 1 \%$), 1.7 GeV ($\approx 6 \%$) and 1.0 GeV ($ 11 \%$) respectively. 

Thus we can say that in the NH and IH cases, 
both base lines have almost equal $\delta_{CP}$ phase 
sensitivity at given oscillation maxima. Although for both NH and IH cases, the two base lines 
have almost equal $\delta_{CP}$ phase sensitivity, but location of given oscillation maximas lies at higher values of beam energies 
in the case of longer base line i.e. LBNO. It is also evident from Fig.~\ref{fig:optmz}, that the gradient of parameter 
$A^M$ w.r.t. the beam energy around peak value of oscillation maxima changes very rapidly
(suggesting very fast oscillations) and this rapidness further increases as we move from first to higher order
maximas. Owing to this reason, we do not prefer to investigate higher oscillation maximas, 
yet these have large sensitivity toward $\delta_{CP}$ variations. Therefore, we can't investigate higher order maximas 
with sharp peaks to the desired precision, in the context of currently available  energy resolutions of the neutrino detectors.

If we look at the shape of the curves in the shaded
region drawn for the spread in beam energy for given experiment, 
curves are almost straight lines. Due to which, we can predict results in terms of average values
over the possible beam energy spreads. We can find from Fig.~\ref{fig:optmz}, that  
$\langle A^M \rangle \approx 2 \%$ (NH-case); $\langle A^M \rangle \approx 1 \%$ (IH-case) at $\langle E \rangle \approx ~3.6 ~GeV $ for LBNE and  
$\langle A^M \rangle \approx 4 \%$ (NH-case);  $\langle A^M \rangle \approx 2 \%$ (IH-case) at 
$\langle E \rangle \approx ~5.0~ GeV$ for LBNO. We can conclude, that there is observable sensitivity towards the 
$\delta_{CP}$ phase variations for both experimental configurations, but to achieve more 
sensitivity towards the variation of $\delta_{CP}$ phase and high precision 
in constraining $\delta_{CP}$ phase, we need to explore observable around higher maximas, 
which can be realized only with nearly mono-energetic beam. 

  Since in accelerator beam sources both $\overline{\nu}_{\mu}$ and $\nu_{\mu}$ beams are 
   equally available, hence, it is possible to study $\overline{\nu}_{\mu}\rightarrow \overline{\nu}_{e}$
    channel experimentally. In case of anti-neutrino, Fig.~\ref{fig:optmz} can be replotted by replacing the NH curves
       with IH ones and vice versa. It is evident from Eq. (\ref{seri}), when we transform from NH-case to IH-case, 
         parameters $\alpha$, $\Delta$ and A change sign, and in the case when we transform from particle to antiparticle, only
          parameters A and $\delta_{CP}$ change sign. If we compare the final results of above two transformations,
        we can find that it is third term that changes sign, while first two terms appear with same sign in the expressions
        for two transformations. We can also find that in the chosen beam energy ranges, shown by shaded regions 
        for the two experiments, the contribution of third term is negligible in comparison to sum of first two terms.
\begin{figure*} 
\caption{(Color online)   An oscillogram of transition probability CP-violation 
         phase sensitivity parameter $A^M$, in the $E$--$L$ plane. Central red dot 
          corresponds to $\langle E \rangle$ and error bar to $\Delta E$ tabulated in table \ref{tab:lblcon}.
          The remaining oscillation parameters have the best fit values shown
in table \ref{tab:bfit}.}
\includegraphics[width=1.0\textwidth]{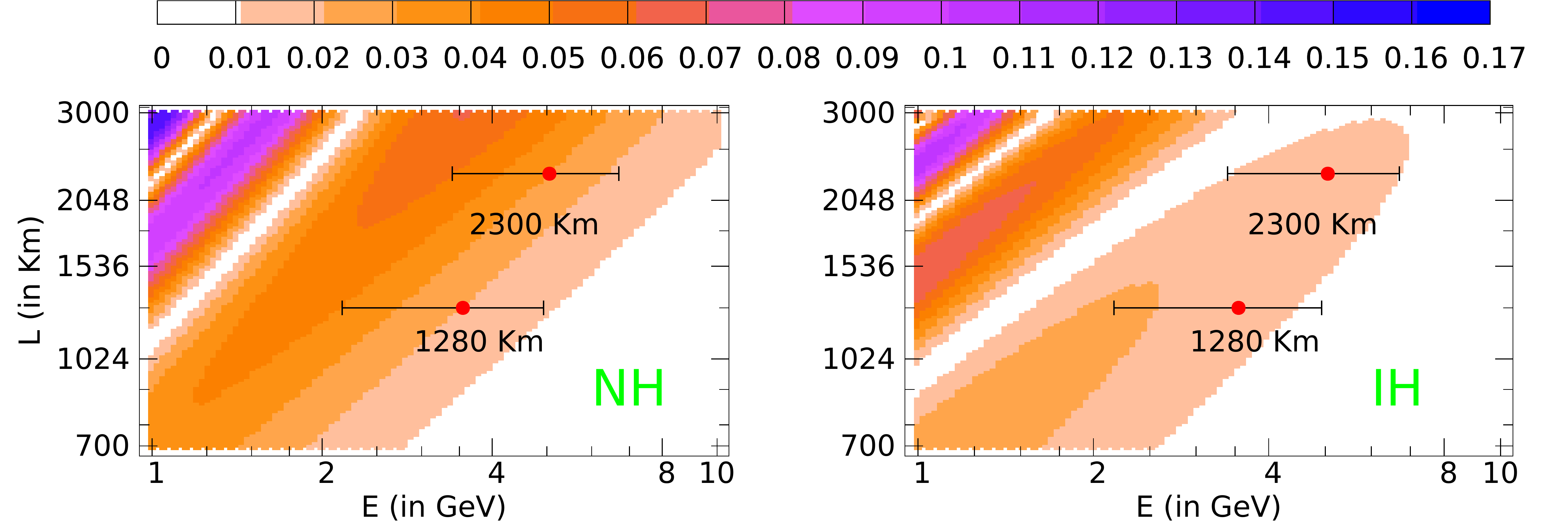}
\label{fig:prcnt}
\end{figure*}
In Fig.~\ref{fig:prcnt}, an oscillogram for the parameter $A^M$ in the $E$--$L$ plane is plotted. 
It is evident from this figure, that for LBNE in the NH-case, average value (i.e. at central red dot 
corresponding to average beam energy) of $A^M$ is $\simeq 2$\%, while for IH-case, it is $\simeq 1.5$\%. 
In the LBNO experiment in the NH-case, the parameter $A^M$ assumes average value of $\backsimeq 3\%$ while
in IH-case, it has value $\backsimeq 1.5\%$. Hence, in case of both experiments sensitivity 
toward the variations in $\delta_{CP}$ phase for the NH-case 
is more as compared to the IH-case. Also, this sensitivity further increases at lower end of energy
spectrum in case of NH and remains almost same over the whole range in the energy spread for IH-case. 
It is recommended to investigate the $\delta_{CP}$ phase at lower values of energy spectrum, 
especially for the confirmed NH-case.
%
%
\setlength{\arrayrulewidth}{0.27 mm}  
\setlength{\tabcolsep}{7pt} 
\renewcommand{\arraystretch}{1.3} 
\begin{table*}[htbp!]
\centering
\caption{The Long Base Line (LBL) experimental configurations \cite{tohln}, considered in the present work.
\label{tab:lblcon}}
\begin{tabular}{ p{3cm} p{3cm} p{3cm} }  \hline
\head{Experiment} &   \head{Baseline} & \head{Beam Energy}  \\ 
   & $L ~(km)$ & $\langle E \rangle \pm \Delta E$ (GeV) \\
\hline
LBNE (DUNE) & 1280 & 3.55 $\pm$ 1.38 \\ 
LBNO & 2300 & 5.05 $\pm$ 1.65  \\
\hline
\end{tabular}
\end{table*}
%
 
                   \section{CP-violation probability, {{\bf \fontsize{16}{5}\selectfont   $\delta_{CP}$ }} 
                     phase sensitivity parameter {{\bf \fontsize{16}{5}\selectfont ($A^{CP}$)}} }
                             \label{sec:cpproptimiz}

We can write an expected event rate at detector site in the following way \cite{ent1,ent1a,ent2,ent3}  
\begin{eqnarray}
 N \backsimeq \langle \phi ~ P ( \nu_{\alpha}\rightarrow \nu_{\beta} ) ~ \sigma ( \nu_{\beta} \rightarrow 
               \beta ) \rangle   \label{evnt}
\end{eqnarray}

where angular bracket denotes the average over neutrino beam energy ($E_{\nu}$), $\phi$ is the neutrino flux at detector site 
and $\sigma$ is the neutrino-nucleon interaction cross section.

Event rate for neutrino and anti-neutrino case from Eq. (\ref{evnt}) can be written as
\begin{eqnarray}
  N_{\nu} \backsimeq \langle \phi_{\nu} ~ P ( \nu_{\alpha}\rightarrow \nu_{\beta} ) ~ \sigma_{\nu} \rangle ~~~~~ and  ~~~~~  
  N_{\overline{\nu}} \backsimeq \langle \phi_{\overline{\nu}} ~ P ( \overline{\nu}_{\alpha}\rightarrow \overline{\nu}_{\beta} ) 
  ~ \sigma_{\overline{\nu}} \rangle                
               \label{evnt1}
\end{eqnarray}
If we consider the case of nearly mono-energetic neutrino beam, which is true for certain off axis beam and 
that both the neutrino and anti-neutrino beam fluxes are nearly equal (i.e. $\phi_{\nu}\backsimeq \phi_{\overline{\nu}} = \phi$),
then we can write
\begin{eqnarray}
 \Delta N^{CP} &=& N_{\nu} - N_{\overline{\nu}} = \phi ~\sigma ~(2~ P_{\alpha \beta} - P_{\overline{\alpha} \overline{\beta}})  \nonumber \\
               &\propto& 2~ P_{\alpha \beta} - P_{\overline{\alpha} \overline{\beta}} = A^{CP} (say)
 \label{evnt2}
\end{eqnarray}

where the fact that $\sigma_{\overline{\nu}} \backsimeq \sigma_{\nu} / 2 = \sigma$,
\cite{ent1,ent1a,crx1,crx2} has been used in the above equation.

We can estimate parameter $A^{CP}$ in case of platinum channel ($\nu_{\mu} \rightarrow \nu_{e}$)
with the help of Eqs. (\ref{seri}) and (\ref{antseri}) to the final form as:
\begin{subequations}
\begin{eqnarray}
A^{CP} &=& 2~ P_{\mu e}(\delta_{CP}) - P_{\bar{\mu} \bar{e}}
(\delta_{CP})  \nonumber \\
 &=& \alpha^2  sin^2{2\theta_{12}} ~c^2_{23} ~\frac{sin^2~{[A \Delta \frac{L}{2}]}}{A^2} + 4~ s^{2}_{13}~ s^{2}_{23} 
 ~\left[\frac{2 ~sin^{2}{[(A-1)\Delta \frac{L}{2}]}}{(A-1)^{2}} -\frac{sin^{2}{[(A+1)\Delta 
 \frac{L}{2}]}}{(A+1)^{2}}\right]    \nonumber  \\
&&  + ~ 2 ~\alpha~ s_{13} ~sin {2 \theta_{12}} ~sin {2\theta_{23}}~\frac{sin~{[A \Delta \frac{L}{2}]}}{A} 
    \left[2 ~cos{(\Delta \frac{L}{2} + \delta_{CP})} ~ \frac{sin{[(A-1) \Delta} \frac{L}{2}]}{(A-1)}  \right. \nonumber \\     
 &&     \left. - ~cos{(\Delta \frac{L}{2} - \delta_{CP})} ~ \frac{sin{[(A+1) \Delta} 
   \frac{L}{2}]}{(A+1)}\right]      \nonumber \\
  & = & g ~ + ~ r1 ~ cos{~\delta_{CP}} ~ + ~ r2 ~ sin{~\delta_{CP}}    
        \label{cpprb}
\end{eqnarray}

where g comprises the first two terms independent of CP-violation phase $\delta_{CP}$ 
and coefficients of the other $\delta_{CP}$ dependent terms have the 
expressions as:  
\begin{eqnarray}
 r1 &=& 2 ~\alpha~ s_{13} ~sin {2 \theta_{12}} ~sin {2\theta_{23}}~\frac{sin~{[A \Delta \frac{L}{2}]}}{A} 
         \left( \frac{2 ~sin[(A-1)\Delta L/2]}{A-1} - \frac{sin[(A+1)\Delta L/2]}{A+1}\right) cos(\Delta L/2)  \nonumber   \\
 r2 &=& -2 ~\alpha~ s_{13} ~sin {2 \theta_{12}} ~sin {2\theta_{23}}~\frac{sin~{[A \Delta \frac{L}{2}]}}{A} 
 \left( \frac{2 ~sin[(A-1)\Delta L/2]}{A-1} + \frac{sin[(A+1)\Delta L/2]}{A+1}\right) sin(\Delta L/2) \nonumber \\
 \label{coeffcp}
\end{eqnarray}

 This parameter enables the measurement of CP-violation phase as long as the constant matter density approximation holds 
  very well. Matter effects along with increasing the oscillation amplitude also increase the 
  sensitivity toward the $\delta_{CP}$ phase variations. 
  Above Eq. (\ref{cpprb}) can be further compacted to the new form, in the following way
\begin{eqnarray}
 A^{CP} = g ~ + ~ \sqrt{r1^{2} ~ + ~ r2^{2}} ~ sin(\gamma + \delta_{CP})    
 \label{prcpm}
\end{eqnarray}
\end{subequations}

where $\gamma = tan^{-1}(r1/r2)$.  
 
The maximum possible $\delta_{CP}$ phase sensitivity of the above CP-violation probability parameter at given beam energy `$E$' and
base line `$L$' can be written as:
\begin{eqnarray}
A_{m}^{CP}(say) \equiv A_{max}^{CP}(\delta_{CP}) -  A_{min}^{CP}(\delta_{CP}) = 2 ~\sqrt{r1^{2} ~ + ~ r2^{2}} 
 \label{cpmax}
\end{eqnarray}

this parameter helps to find an optimal beam energy for given base line 
and the optimal experimental base line for given beam energy, for which $\delta_{CP}$ phase
sensitivity is maximum. 
\begin{figure*}[htbp!]
\caption{(Color online). Variation of the working parameters; $A^{CP}$ 
defined in Eq. (\ref{prcpm}) [shown by magenta colored band
for 0--$2 \pi$ variation in the $\delta_{CP}$-phase] and maximum possible 
$\delta_{CP}$ phase sensitivity parameter i.e. $A^{CP}_m$ (band width) calculated in Eq. (\ref{cpmax}) 
[shown by black curve], as a function of the beam energy $E$. Thus $A^{CP}_m$ is a measure of 
band width in the broad curve.  The remaining oscillation parameters have the best fit values shown
in table \ref{tab:bfit}.}
\includegraphics[width=0.95\textwidth]{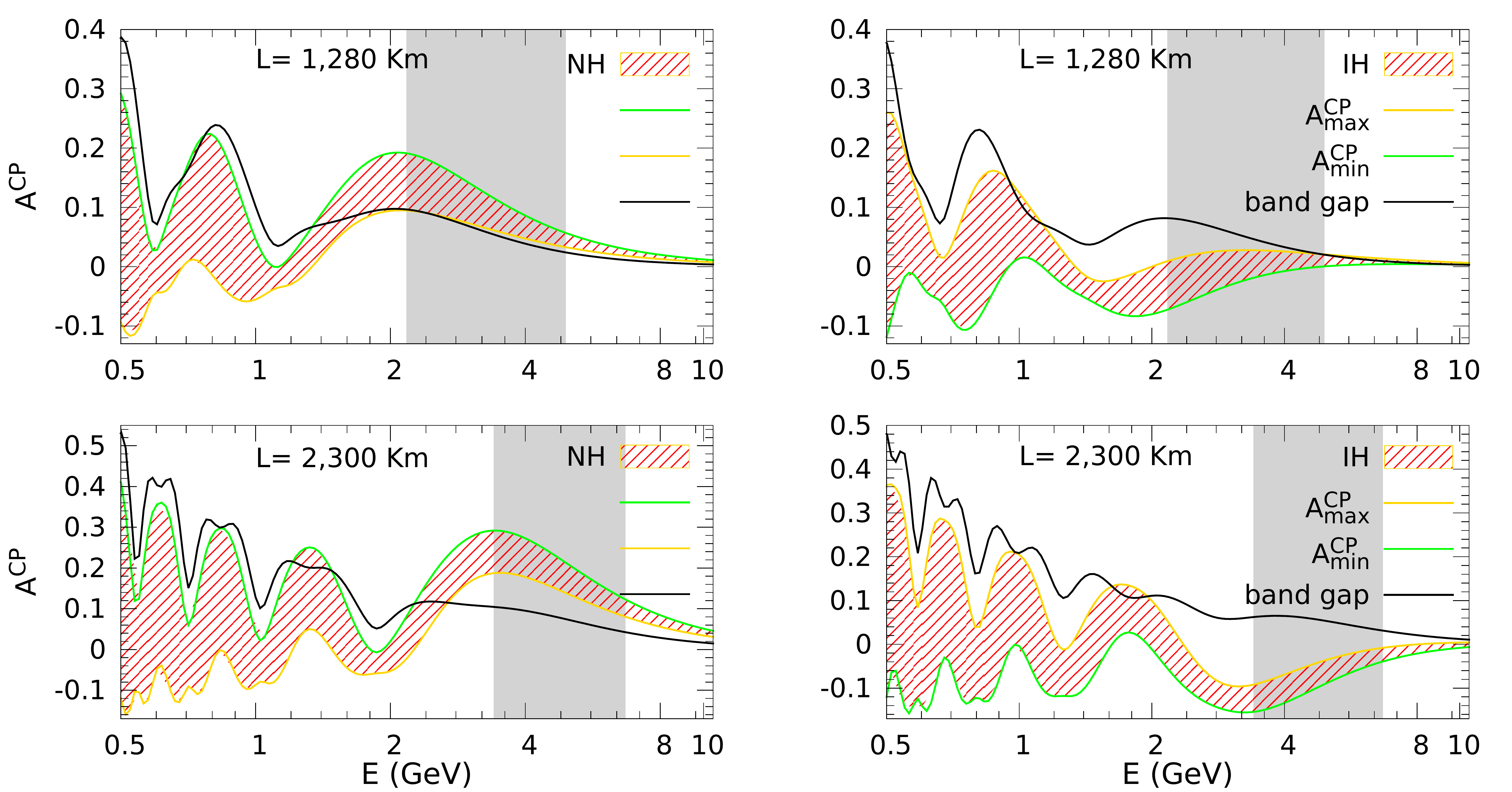}
\label{fig:dcpNI}
\end{figure*}
 This CP-violation probability, $\delta_{CP}$ phase sensitivity parameter $A^{CP}$ (for $\delta_{CP} \rightarrow 0, ~2 ~\pi$)
 in case of both the NH and IH cases, is illustrated as a function of beam energy $E$ in Fig.~\ref{fig:dcpNI}, for the chosen LBL experimental 
 setups viz. LBNE \& LBNO. 
 
  We will restrict our discussion mainly to and nearby the first oscillation maxima, i.e. 
    for $E>1$ GeV in case of LBNE and $E> 2$ GeV in the LBNO case. In these figures we observe that in both
    the NH and IH cases, for LBNE we expect an average sensitivity of $\approx 3 \%$ at $\langle E \rangle \approx 3.55 ~GeV$ 
    and for LBNO there is a sensitivity of $\simeq 6\%$ at $\langle E \rangle \approx 5.05 ~GeV$.
   
   There are other oscillation maximas, for example, at $E\approx0.8$ GeV with sensitivity of $\approx 25 \%$ for LBNE
   in both the NH and IH cases. While for the LBNO experiment at E=1.3 GeV we expect a sensitivity of $\approx 20 \%$ 
   for NH-case and sensitivity of $\simeq 12 \%$ for IH-case. Also, there are other oscillation maximas with 
   sensitivity of $\simeq 32, ~42$ \% for the NH-case and  $\simeq 20, ~30$ \% for IH-case at 0.8, 0.6 GeV respectively. 
   But due to fast oscillations around these maximas, almost mono-energetic beam energy could only make 
      the experimental realization possible. Energy spreads in the currently available beam sources are relatively broad, due to which
     we don't prefer to discuss about these oscillation maximas in detail. 

  The other thing we notice in these figures is that, in the specific beam energy range,
  parameter $A^{CP}$ attains positive values for one hierarchy and negative 
  for the other hierarchy. For example, in the case of LBNO experiment, in the beam energy range of 
 $2-8 ~GeV$, parameter $A^{CP}$ assumes positive values in the NH-case, while negative values
  in the IH-case over the whole $\delta_{CP}$ ($0-2 ~\pi$) possible range. 
  These positive and negative values of the parameter $A^{CP}$ can be confirmed experimentally.
  Thus these energy ranges provide the opportunity to investigate mass ordering (MO) along with 
  $\delta_{CP}$ phase investigation.

\begin{figure*} 
\caption{(Color online)  CP violation probability, $\delta_{CP}$
          phase sensitivity parameter ($A^{CP}_m$) oscillogram in the $E-L$ plane. Central red dot 
          corresponds to $\langle E \rangle$ and error bar to $\Delta E$ tabulated in table \ref{tab:lblcon}.
          The remaining oscillation parameters have the best fit values shown in table \ref{tab:bfit}.}
\includegraphics[width=1.0\textwidth]{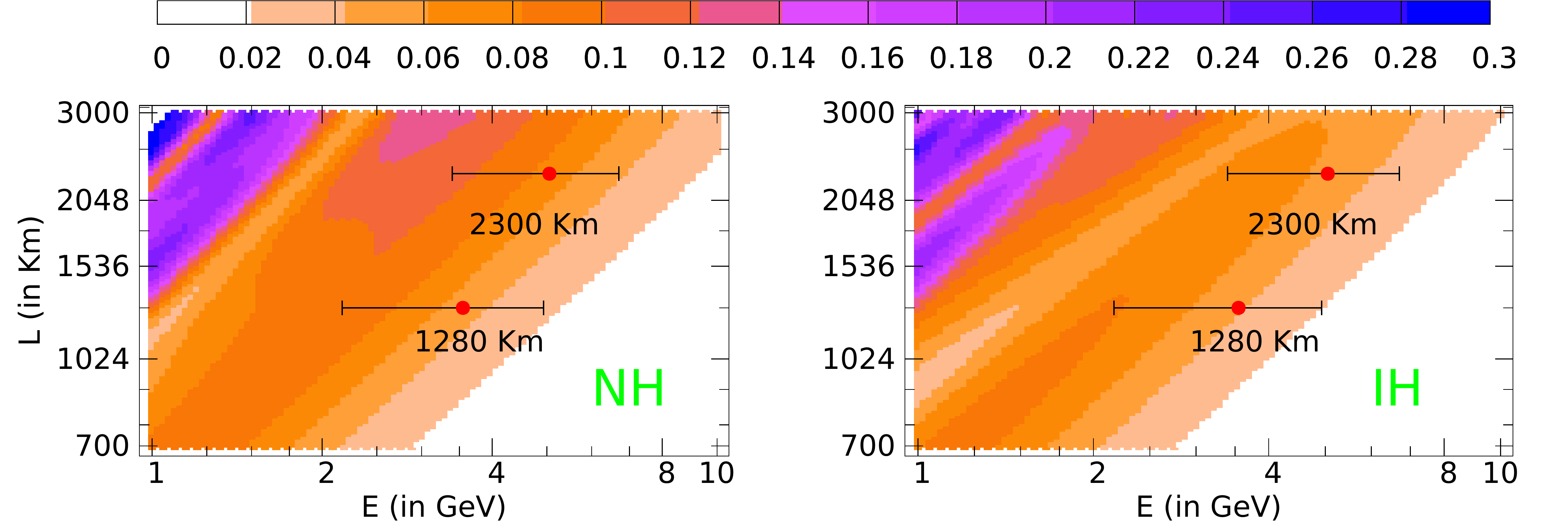}
\label{fig:dcpcnt}
\end{figure*}

In Fig.~\ref{fig:dcpcnt}, an oscillogram for the parameter $A^{CP}_m$ in the $E-L$ plane has been shown.
It is evident from the oscillogram that, at  central red dot for LBNE(DUNE), $A^{CP}_m$ is $\simeq 5 \%$  
for NH-case and $\simeq 4 \%$ in IH-case. At lower end of the energy spectrum we expect 
a sensitivity of $\simeq 9 \%$, while at higher end of $\simeq 2 \%$ for both hierarchies.
In the LBNO experiment, in the NH-case, at central red dot $A^{CP}_m \simeq 8 \%$  
and further has values of $ \simeq 11 \%$  \& $\simeq 4 \%$  respectively at lower \& higher
ends of energy spectrum. While in the IH-case, 
parameter $A^{CP}_m$ has value $\simeq 5 \%$ at
central red dot and has values of 
$\simeq 7 \%$  \& $\simeq 3 \%$  
at lower \& higher ends of energy spectrum 
respectively. Thus for LBNO, normal hierarchy has 
more sensitivity as compared to inverted hierarchy and 
the sensitivity further increases at lower end of energy spectrum. It suggests that 
lower energy spectrum ends are more suitable than higher ones to investigate $\delta_{CP}$ phase.

                   \section{Sensitivity of {{\bf \fontsize{16}{5}\selectfont $A^M$}} \& {{\bf \fontsize{16}{5}\selectfont
                    $A^{CP}$}} towards mixing angles and mass square differences variations}
                             \label{sec:senstivity}
\begin{figure*}[htbp!]
\caption{(Color online) The sensitivity of parameter $A^{CP}_m$ within 3$\sigma$ variations
of mixing angles and mass square differences, for experimental configurations LBNE ($E=3.55 ~GeV, ~L=1280 ~km$,  
$\rho=3 g/cm^3$) [in first two columns] and LBNO ($E=5.05 ~GeV, ~L=2300 ~km$, $\rho=3.3 g/cm^3$) [in the third and fourth columns].
The parameter $A^{CP}_m \equiv A^{CP}_m (\mu  \rightarrow e)$ along the 
y-axis is in the units of $10^{-2}$. All the other oscillation parameters, except the one considered along x-axis, assume the best fit 
values, as tabulated in table \ref{tab:bfit}.}
\includegraphics[width=0.495\textwidth]{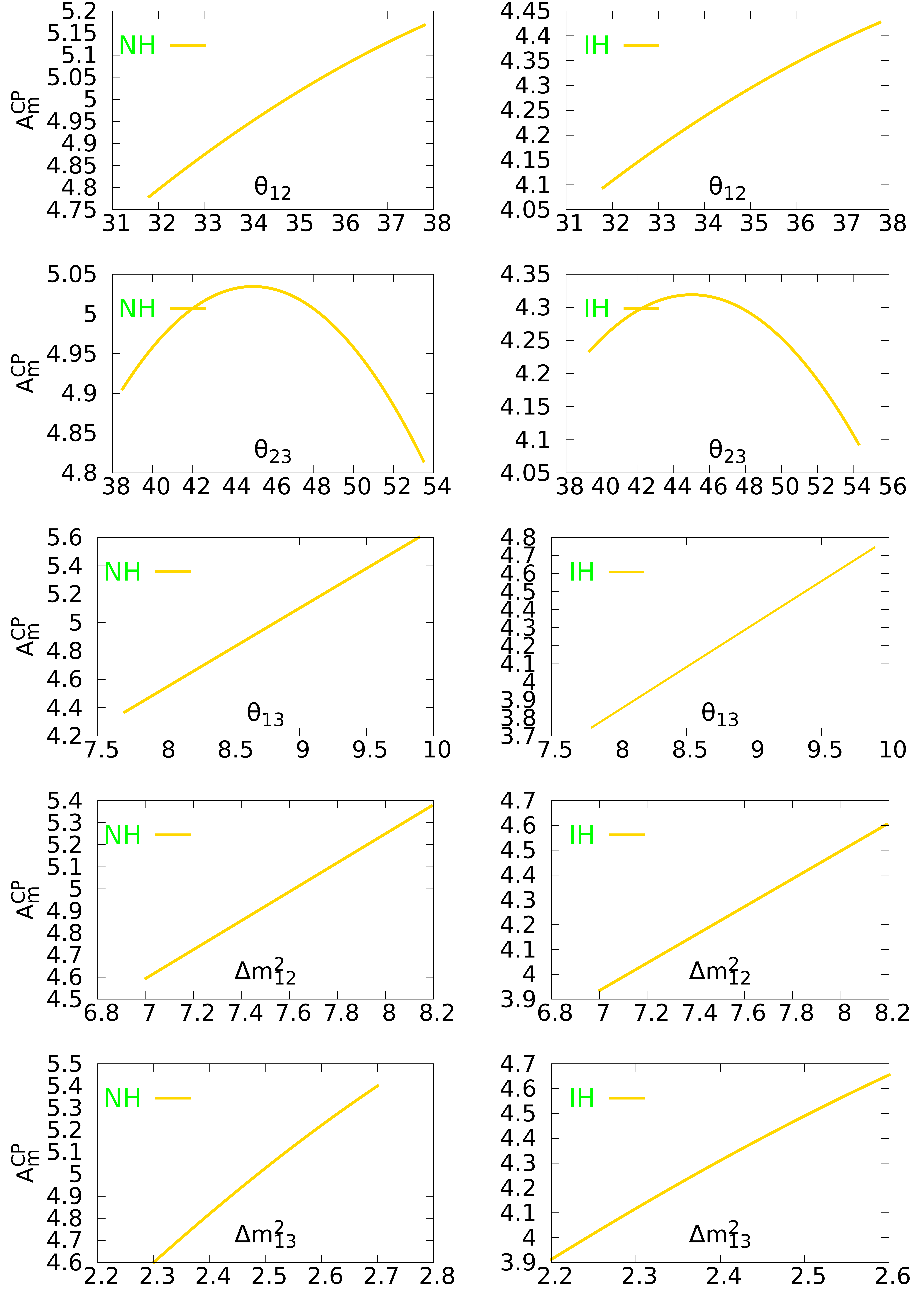}
\includegraphics[width=0.495\textwidth]{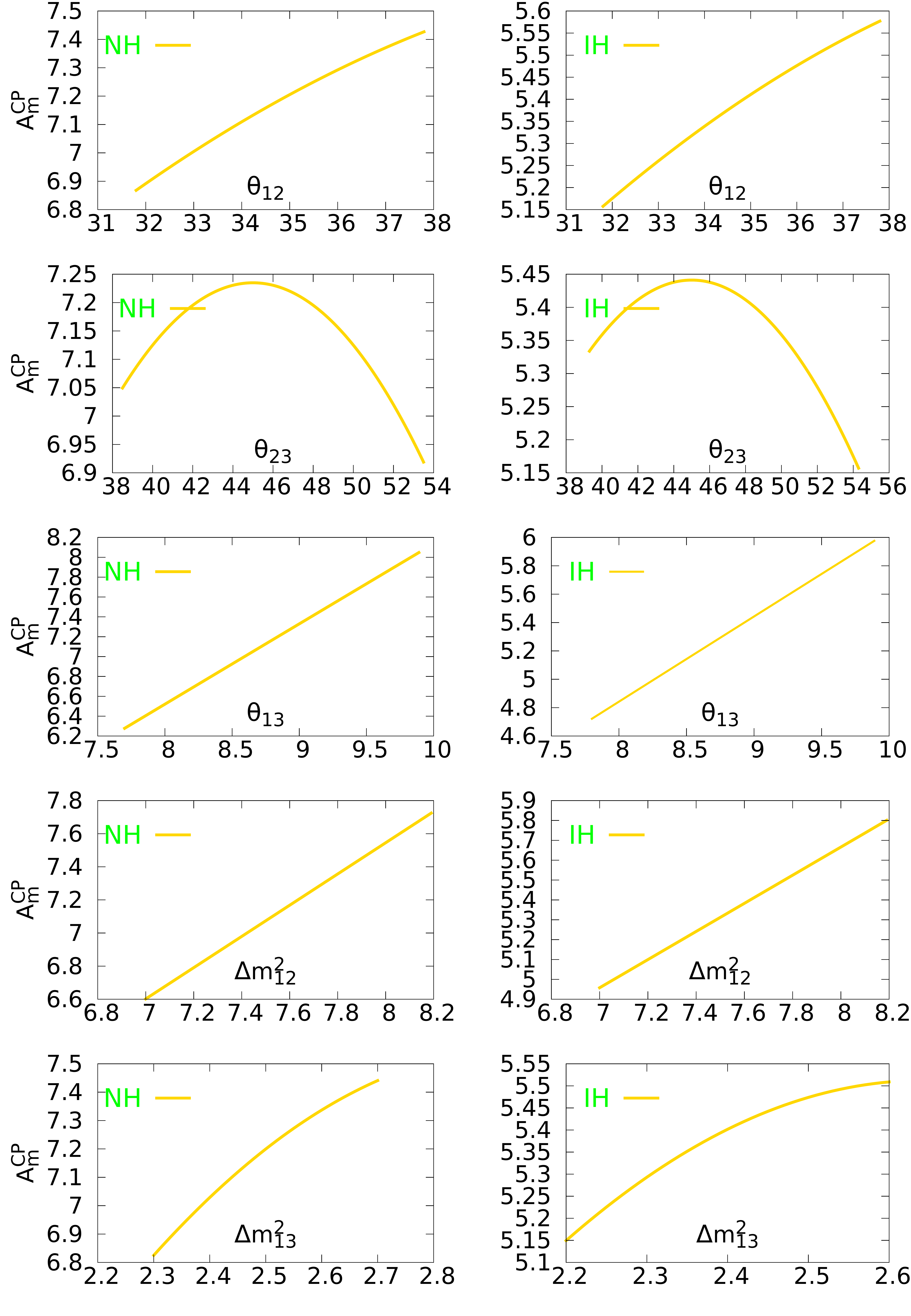}
\label{fig:prasens}
\end{figure*}

\begin{figure*}[htbp!]
\caption{(Color online)   The sensitivity of parameters $A^{CP}$ and $A^M$ within $3 \sigma$ variations
of mixing angles and mass square differences, for experimental configuration ($E=0.5 ~GeV, ~L=1280 ~km$, $\rho=3 g/cm^3$). 
The parameter $A^{CP}_m \equiv A^{CP}_m (\mu  \rightarrow e)$ along the 
y-axis is in units of $10^{-2}$. Where first column (NH-case) and second column (IH-case) correspond to parameter $A^{CP}$, while third (NH-case)
and fourth (IH-case) columns correspond to the parameter $A^M$.  
All other oscillation parameters, except the one considered along x-axis, assume the best fit 
values, as tabulated in table \ref{tab:bfit}.}
\includegraphics[width=0.495\textwidth]{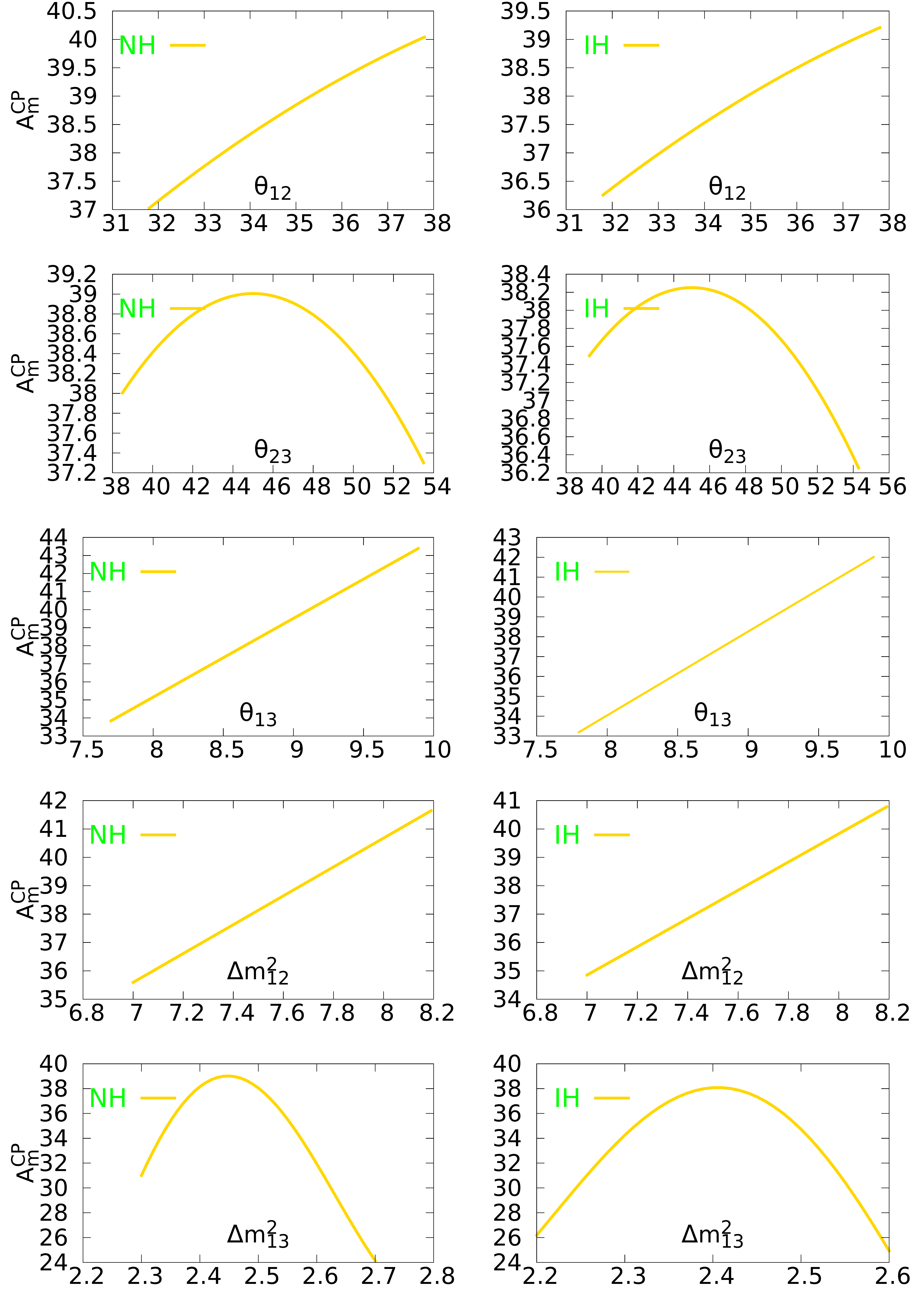}
\includegraphics[width=0.495\textwidth]{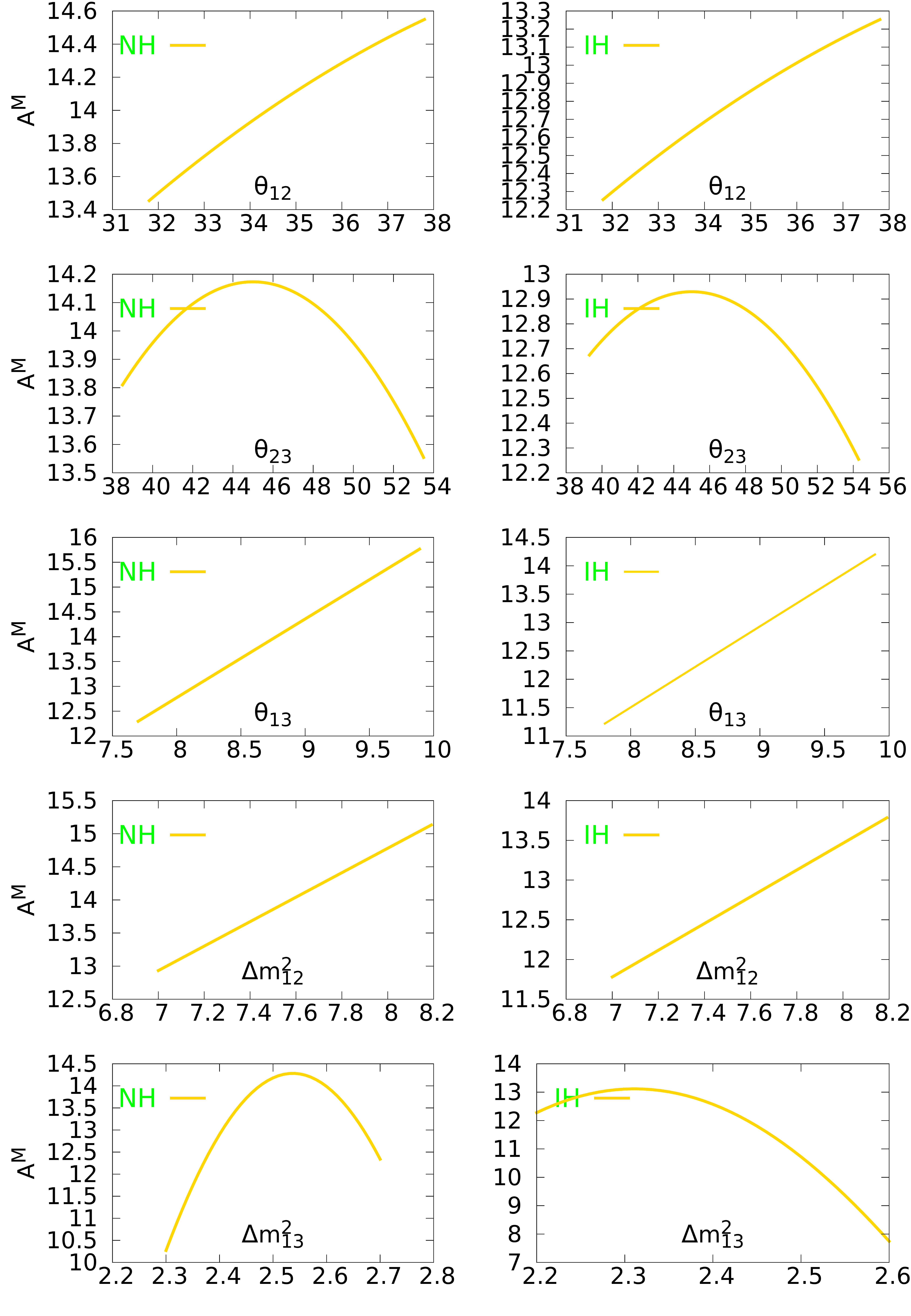}
\label{fig:prasens12}
\end{figure*}

\begin{figure*}[htbp!]
\caption{(Color online)   The sensitivity of parameter $A^M$ and $A^{CP}_m$ within 3$\sigma$ variations
of mixing angles and mass square differences, for experimental configuration ($E=0.5 ~GeV, ~L=2300 Km$, $\rho=3.3 g/cm^3$).
The parameter $A^{CP}_m \equiv A^{CP}_m (\mu  \rightarrow e)$ along the 
y-axis is in units of $10^{-2}$. Where first column (NH-case) and second column (IH-case) correspond to parameter $A^{CP}_m$, 
while third (NH-case) and fourth (IH-case) columns correspond to parameter $A^M$.  
All the other oscillation parameters, except the one considered along x-axis, assume the best fit 
values, as tabulated in table \ref{tab:bfit}.}
\includegraphics[width=0.495\textwidth]{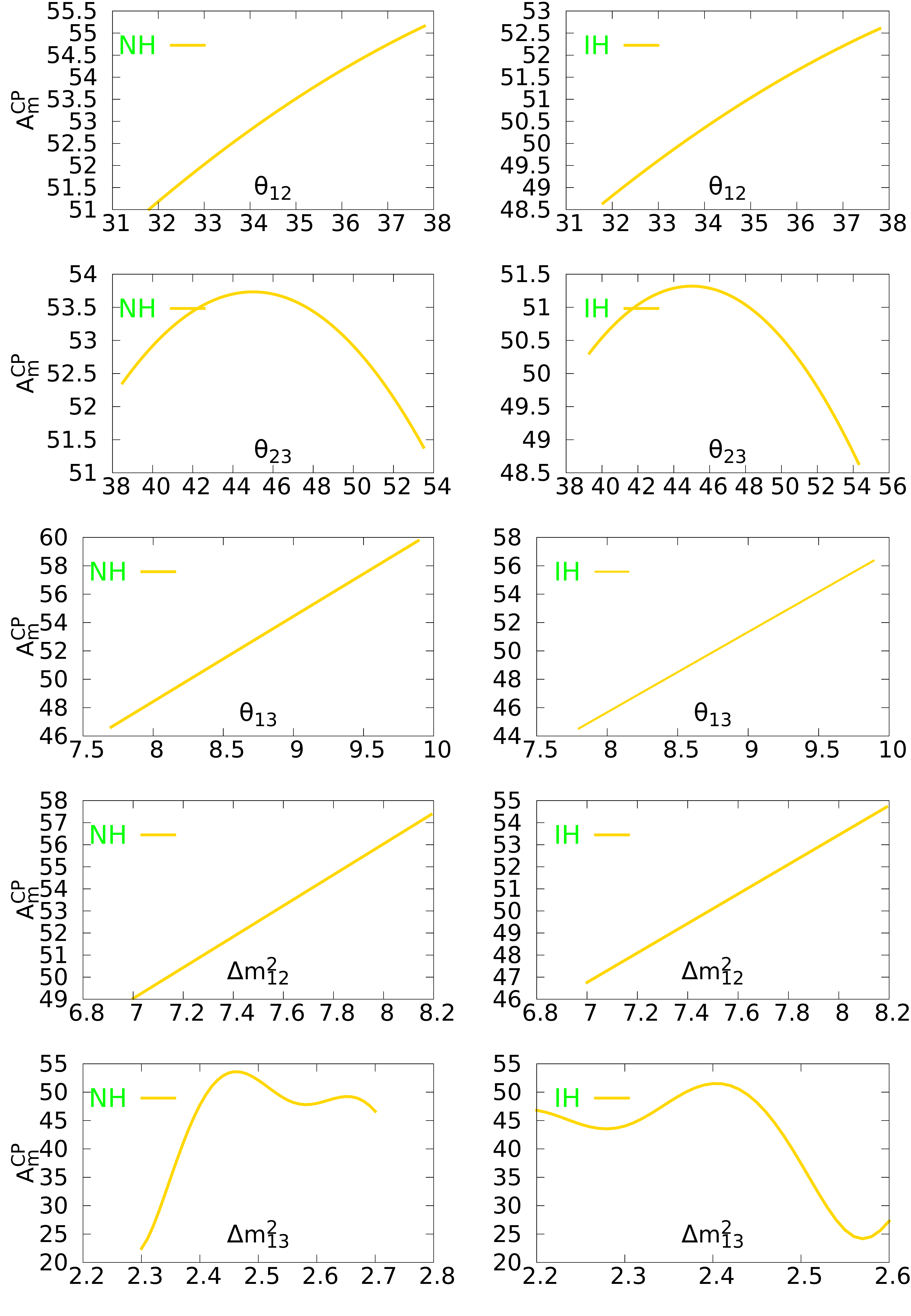}
\includegraphics[width=0.495\textwidth]{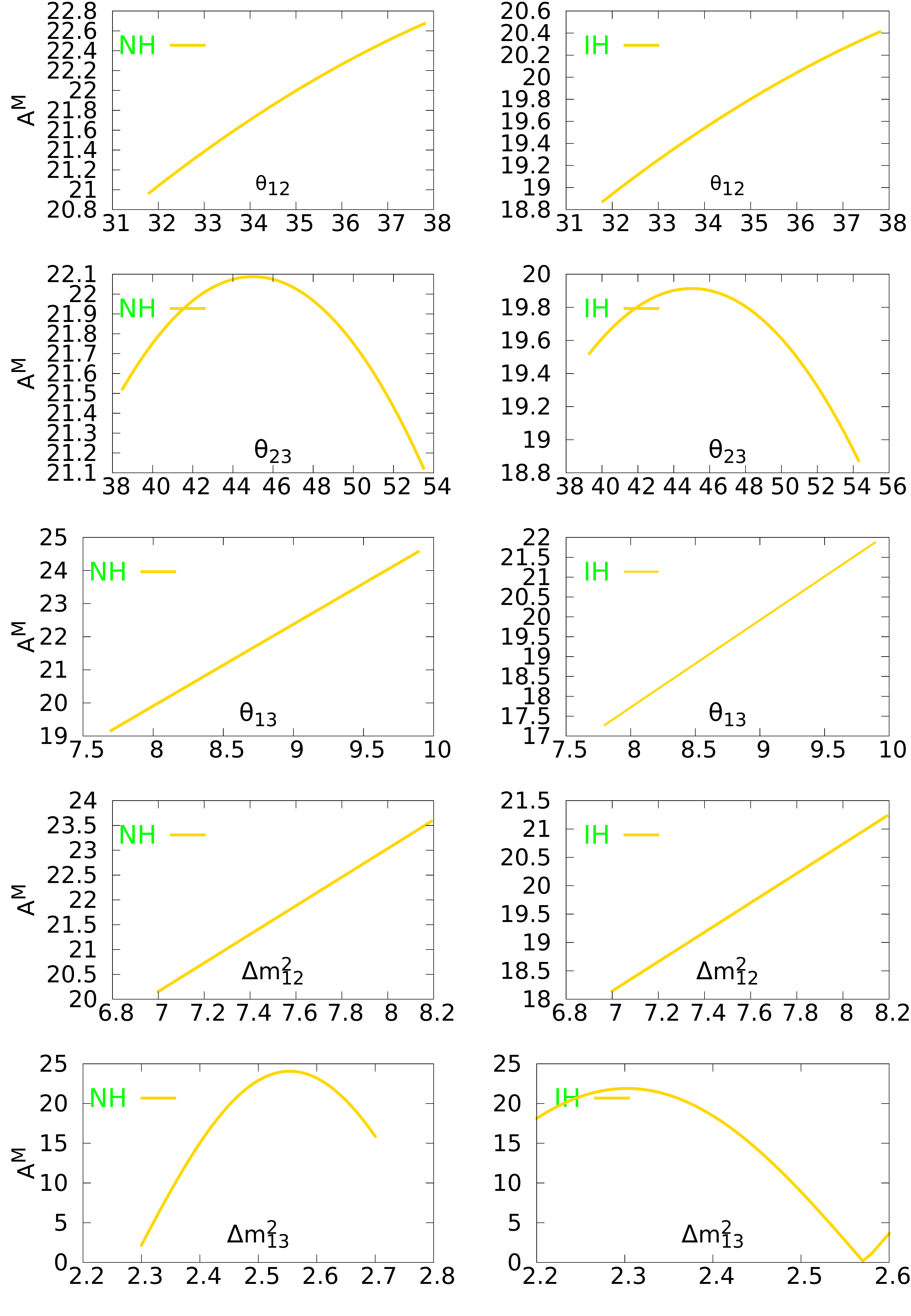}
\label{fig:prasens23}
\end{figure*}

In Figs.~\ref{fig:prasens}, \ref{fig:prasens12} and \ref{fig:prasens23} sensitivity of working parameters 
$A^{CP}_m$ and $A^M$ toward the $3 \sigma$ variations in three mixing angles and two 
mass square differences is illustrated. It is clear from Fig.~\ref{fig:prasens}, that parameter 
$A^{CP}_m$ is feebly sensitive towards the variations in mixing parameters in the chosen range, for
both LBNE and LBNO experiments. From the numerical analysis, it can also be confirmed that parameter $A^M$ 
has also small sensitivity to the variations in mixing parameters for both the experimental configurations.


In Fig.~\ref{fig:prasens12}, sensitivity of the parameters $A^{CP}_m$ and $A^M$ towards the variations in mixing
parameters for experimental configuration ($E=0.5 ~GeV, ~L=1280 ~km$) has been illustrated. We can 
analyze from the figure, that both parameters have small sensitivity towards mixing angles 
$\theta_{12}$ \& $\theta_{23}$ variations. Parameter $A^{CP}_m$ has noticeable sensitivity towards the 
$\theta_{13}$ variations, while $A^M$ has small sensitivity. Also, the parameter $A^{CP}_m$ has noticeable sensitivity
to the $\Delta m^2_{12}$ variations, while parameter $A^M$ has small sensitivity.
We can easily notice that $A^{CP}_m$ has large sensitivity towards the $\Delta m^2_{13}$ variations, while 
$A^M$ has moderate sensitivity on that parameter.
We can conclude that, if we know $\theta_{13}$ very precisely, then parameter $A^{CP}_m$ is left sensitive 
to the mass square differences.

It is evident from the analysis of Fig.~\ref{fig:prasens23} that, parameter $A^{CP}_m$ attains moderate 
sensitivity towards the $\theta_{12}$ and $\theta_{23}$ variations, while $A^M$ has small sensitivity towards these variations.
Also parameter $A^{CP}_m$ has large sensitivity to the $\theta_{13}$ variations, while parameter 
$A^M$ has moderate sensitivity. Similarly $A^{CP}_m$ has noticeable sensitivity towards 
the $\Delta m^2_{12}$ variations, while $A^M$ has moderate sensitivity. 
It is evident from figure, that both parameters $A^{CP}_m$ and $A^M$ attain large sensitivity
towards the variations in the atmospheric mass square difference ($\Delta m^2_{13}$). Thus if we know the precise value 
of reactor mixing angle $\theta_{13}$ then, we are left with large sensitivities of the 
parameters toward the variations in solar and atmospheric mass square differences only. 
Parameter $A^{CP}_m$ has large sensitivity towards the solar mass square difference (i.e. $\Delta m^2_{12}$) variations in 
comparison to parameter $A^M$, while both $A^{CP}_m$ \& $A^M$ have large sensitivity towards the variation in atmospheric mass square difference 
(i.e. $\Delta m^2_{13}$) in the 3$\sigma$ range.

We can also conclude that for given parameter, sensitivity in the NH-case is always almost equal to  
sensitivity in IH-case. It is also clear from the analysis of all the three Figs. (i.e. \ref{fig:prasens}, 
\ref{fig:prasens12}, \ref{fig:prasens23}), that the sensitivity of parameters increase with the increase in 
baseline length `$L$' and lowering in the beam energy `$E$'.

If we compare parameters $A^{CP}_m$ and $A^M$, then former shows large sensitivity in comparison to 
latter.

                                    \section{Conclusions and perspectives}
                                     \label{sec:conclusion}

In this work we have studied two parameters viz. the `` transition probability, $\delta_{CP}$ phase sensitivity parameter, $A^M$ '' 
and the `` CP-violation probability, $\delta_{CP}$ phase sensitivity parameter, $A^{CP}$ ''   especially to investigate Dirac's $\delta_{CP}$  
phase. We can conclude from the analysis of Figs.~\ref{fig:optmz}, \ref{fig:prcnt}, \ref{fig:dcpNI} and \ref{fig:dcpcnt}, that
LBNO provides better sensitivity as compared to LBNE towards the $\delta_{CP}$ phase variations. We also notice that for a given
baseline, this sensitivity is more in the NH-case than in the IH-case, for the parameter $A^M$. This sensitivity further increases at
lower end of energy spectrum. Parameter $A^{CP}$ enables to clearly differentiate among the two mass hierarchies from the 
positive and negative values of this parameter for either of hierarchy, in the specific beam energy range,
as is apparent from Fig.~\ref{fig:dcpNI}. This latter distinction of two hierarchies by the sign of $A^{CP}$ is more pronounced
in case of LBNO experiment. We also notice from the comparison of Figs.~\ref{fig:prcnt} and \ref{fig:dcpcnt}, that for given 
base line length ($L$) and beam energy ($E$) sensitivity towards the $\delta_{CP}$ 
phase variations is larger for the parameter $A^{CP}_m$ in comparison to parameter $A^M$.

Though it is not convenient to investigate higher oscillation maximas (i.e., second and third order), due to 
large spreads in the energy spectrum of beam sources and comparatively less resolutions of the present detectors.
But, sufficiently large values of parameters $A^M$ and $A^{CP}_m$ at these maximas encourage to investigate these experimentally,
to put more stringent bounds on the $\delta_{CP}$ phase. Investigation of these 
higher oscillation maximas need more accurate detection techniques and almost narrow energy spectra.

At relatively small value of beam energy, `$E$' ($\simeq 0.5 ~GeV$), sensitivity of the parameter $A^{CP}_m$ 
is large towards the atmospheric mass square difference ($\Delta m^2_{13}$) variations in case of both LBNE and LBNO experiments.
While, this parameter has  moderate sensitivity towards the solar mass square difference ($\Delta m^2_{12}$) variations, in the LBNE experiment case 
and has large sensitivity in the case of LBNO. Parameter $A^M$ has moderate sensitivity towards the $\Delta m^2_{13}$ 
variations and has small sensitivity to the $\Delta m^2_{12}$ variations in case of LBNE, while 
this parameter has large sensitivity to $\Delta m^2_{13}$ variations and has moderate 
sensitivity to the $\Delta m^2_{12}$ variations in the case of LBNO experiment. Also parameter
$A^{CP}_m$ has large sensitivity towards the $\theta_{13}$ variations in case of both LBNE and LBNO
experiments, while the parameter $A^M$ has moderate sensitivity. In case of both LBNE and LBNO
experiments sensitivity of both parameters is small toward the $\theta_{12}$ and $\theta_{23}$
variations. Thus if we know precise value 
of reactor mixing angle $\theta_{13}$, we are left with large sensitivities of the 
parameters toward the variations in solar and atmospheric mass square differences only.
If we need large/small sensitivities, we can prefer to low/high beam energy regions.

As parameters $A^{CP}$ \& $A^{CP}_m$ are the differences of two CP conjugate channels and parameter $A^M$ is that 
of single oscillation channel, no doubt errors/uncertainties get canceled to an extent for both parameters,
but being a difference involving the same channel such cancellation is large in the case of parameter $A^M$. \\

\section*{Acknowledgments}
 I would like to thank Prof. Brajesh Chandra Choudhary (Department of Physics \& Astrophysics, Delhi University)
 for useful discussions.

                   
\pagebreak

\end{document}